\documentclass[fleqn,usenatbib]{mnras}

\usepackage{newtxtext,newtxmath}

\usepackage[T1]{fontenc}

\DeclareRobustCommand{\VAN}[3]{#2}
\let\VANthebibliography\thebibliography
\def\thebibliography{\DeclareRobustCommand{\VAN}[3]{##3}\VANthebibliography}
\usepackage{amstext}
\usepackage{url}
\usepackage[]{times,refname,amsmath,bm}
\bibpunct{(}{)}{;}{a}{}{,}
\usepackage{tabularx}
\usepackage{graphicx,epsfig,color,latexsym}
\renewcommand{\d}{\mathrm{d}}

\newcommand{\bea}{\begin{eqnarray}}
\newcommand{\eea}{\end{eqnarray}}
\newcommand{\be}{\begin{equation}}
\newcommand{\ee}{\end{equation}}
\newcommand{\rund}[1]{\left(#1\right)}

\newcommand{\eck}[1]{\left[ #1 \right]}

\title[B-field in plasma lens]{Plasma lensing with magnetic field and a small correction to the Faraday rotation measurement}

\author[Er et al.]{Xinzhong Er$^{1}$\thanks{Email: phioen@163.com}, Ue-Li Pen$^{2,3,4,5,6}$, Xiaohui Sun$^7$, Dongzi Li$^8$ \\
$^1$ South-Western Institute for Astronomy Research, Yunnan University, Kunming, 650500 P.R.China\\
$^2$ Institute of Astronomy and Astrophysics, Academia Sinica, Astronomy-Mathematics Building, No. 1, Section 4, Roosevelt Road, Taipei 10617, Taiwan\\
$^3$ Canadian Institute for Theoretical Astrophysics, University of Toronto, 60 St. George Street, Toronto, ON M5S 3H8, Canada\\
$^4$ Dunlap Institute for Astronomy $\&$ Astrophysics, University of Toronto, AB 120-50 St. George Street, Toronto, ON M5S 3H4, Canada\\
$^5$ Canadian Institute for Advanced Research, 661 University Avenue, Toronto, Ontario M5G 1M1, Canada\\
$^6$ Perimeter Institute for Theoretical Physics, 31 Caroline St. North, Waterloo, ON, Canada N2L 2Y5\\
$^7$ School of Physics and Astronomy, Yunnan University, Kunming, 650500 P.R.China\\
$^8$ TAPIR, Walter Burke Institute for Theoretical Physics, Mail Code 350-17, Caltech, Pasadena, CA 91125, USA
}
\pubyear{2023}

\begin{document}
\label{firstpage}
\pagerange{\pageref{firstpage}--\pageref{lastpage}}
\maketitle

\begin{abstract}
Plasma lensing displays interesting characteristics that set it apart from gravitational lensing. The magnetised medium induces birefringence in the two polarisation modes. As the lensing deflection grows stronger, e.g. when images form near the critical curve, the geometric delay of the signal can cause rotation in linear polarisation, in addition to Faraday rotation. This rotation has a frequency dependence to the power of four. We study the geometric rotation of the lensed image in a Gaussian density model and find that it is necessary to take into account the geometric rotation when estimating magnetised media, especially in the under-dense lens. At frequencies of $\sim 1$ GHz or lower, the geometric rotation can dominate.
We simulate the flux of lensed images and find that when the image forms near the lensing critical curve, the birefringence can convert the linear polarisation and un-polarisation pulse into a circular mode. The lensing magnification has the potential to increase the probability of detecting such events.

\end{abstract}


\section{Introduction}
The deflection of light rays caused by the inhomogeneous distributions of free electrons is called plasma lensing \citep[e.g.][]{romani87,CleggFL1998,cordesRickett1998}. Initially, it was proposed to explain the Extreme Scattering Events (ESEs), which are compact radio sources occasionally observed to go through a period of demagnification at low frequencies \citep[][]{ESE0}. Some phenomena in the scintillation of pulsars can be also related to the highly anisotropic scattering of the interstellar medium (ISM) \citep[e.g.][]{stinebring2001,2004MNRAS.354...43W,2006ApJ...637..346C,2016MNRAS.458.2509B,2018ApJ...861..132L,2022arXiv220806884Z}. Plasma lensing shares several similar features with gravitational lensing, especially in terms of the mathematical description \citep[][]{2020arXiv200616263W}. However, different features are also introduced by plasma lensing, such as wavelength dependence, opposite to the deflection due to gravity. Moreover, mass models in gravitational lensing have been extensively studied from both numerical simulations and observations, while the free electron density lacks solid constraints. The Gaussian density profile provides a valuable model to describe the behaviours of a discrete clump of ionised material in the ISM \citep{CleggFL1998,romani87}. Since then the plasma lens model has been expanded on in a number of ways \citep[e.g.][]{2012MNRAS.421L.132P,2014MNRAS.442.3338P,er&rogers18,2018MNRAS.478..983S,2018Natur.557..522M}.

When the radio signal propagates through a magnetic medium, the magnetic field can cause rotation to the linearly polarised signal \citep[e.g.][]{1975JPlPh..13..571I,2010ApJ...718.1085B,2019IJMPD..2840013T}. The pulsar Rotation Measure (RM) has been widely used to study the line-of-sight averaged magnetic field \citep[e.g.][]{2018ApJS..234...11H,2020MNRAS.496.2836N,2022ApJ...940...75D,2022arXiv220507917L}. Moreover, the magnetic field induces birefringent in the plasma lensing \citep[e.g.][]{2019ApJ...870...29S,2019MNRAS.486.2809G,2019MNRAS.484.5723L,2020MNRAS.493.1736R}. The birefringent or the extra delay caused by a magnetic field usually is small enough to be neglected. However, some recent observations suggest large magnetic fields, e.g. the repeating FRB 121102 \citep{2014ApJ...790..101S,2018Natur.553..182M,2018ApJ...868L...4M}, the Galactic central magnetar J1745-2900 \citep{2013Natur.501..391E,2018ApJ...852L..12D}. Moreover, a large variation of RMs suggests a complicated intergalactic and interstellar environment \citep[e.g.][]{2021ApJ...908L..10H,2022Sci...375.1266F,2022arXiv220211112A}. The variations of the density of magnetised media can cause multi-path scattering, widen the pulse or change the polarisation mode of the signal \citep[e.g.][]{2000ApJ...545..798M,2016ApJ...824..113X,2019MNRAS.484.5723L,2022MNRAS.510.4654B,2022arXiv220803332K}.
The large RM and its variations can deflect the light and cause the corresponding lensing effects. In this work, we study plasma lensing due to the magnetised medium, especially the time delay and polarisation modes by the lensing. In section \ref{sec:formulae}, we present the basic equations and the idea of geometric rotation and we show some examples in section \ref{sec:example} and give a summary at the end.


\section{The plasma deflection with magnetic field}
\label{sec:formulae}
The basics of lensing can be found in e.g. \citet{2006glsw.conf....1S,2016ApJ...817..176T}.
We use angular diameter distances between the lens and us, the source and us, and between the lens and the source as $D_d,D_s,D_{ds}$ respectively. We introduce the angular coordinates $\theta$ in the lens plane as image position, and those in the source plane $\beta$. They are related by the lens equation
\be
\beta =\theta-\alpha(\theta)
=\theta - \nabla_\theta \psi(\theta),
\ee
where $\alpha$ is the reduced deflection angle, $\nabla_\theta$ is the gradient on the image plane, $\psi$ is the effective lens potential. All the lensing distortions can be calculated from $\psi$. The magnification produced by a lens is inversely related to the Jacobian $A$ of the lens equation, $\mu^{-1}=$ det$(A)$. In order to calculate the plasma lensing potential, we start with the refractive indices for plasma with a magnetic field \footnote{ In this work, we only consider free electrons for the plasma lens. The ion plasma can slightly change our result but is not taken into account in the current study. See \cite{1975JPlPh..13..571I} for more details.}
\be
n_{L,R} 
\approx 1- \frac{\omega_e^2}{2 \omega^2}\rund{1\pm \frac{\omega_B}{\omega}},
\label{eq:refractive-index}
\ee
where the subscript $L,R$ indicate the left- and right-modes of the circular polarisation respectively. The approximation in Eq.\,\ref{eq:refractive-index} holds if $\omega_B\ll \omega$ and $\omega_e\ll \omega$, where $\omega$ is the observational frequency.
$\omega_e$ is the plasma frequency given by
\be
\omega_e^2 = \frac{4\pi n_e(r) e^2}{m_e},  
\ee
which depends on the electron number density $n_e(r)$ in three dimension, and $r$ is the radius. $\omega_B$ is the cyclotron frequency of the parallel magnetic field
\be
\omega_B = \frac{eB \cos \eta }{m_e c},
\ee
with $\eta$ is the angle between magnetic field $\vec{B}$ and the line of sight. Due to the changing of propagation speed in the plasma, there will be a delay of the signal with respect to that in the vacuum
\be
t_{\psi}=\frac{\lambda^2 r_e}{2 \pi c}DM \pm \frac{\lambda^3}{\pi c} RM.
\ee
The plasma lensing potential can be obtained in analogy with gravitational lensing
\be
\psi=\frac{c D_{ds}}{D_d D_s} t_{\psi}.
\ee
The projected electron density, which can be estimated approximately by Dispersion Measure (DM)
\be
DM (\theta)\approx N_e(\theta)= \int n_e(r)\, d z_l.
\ee
The integral is performed alone the light of sight, $z_l$. The homogeneous plasma outside the lens will not contribute to the deflection angle, but to the time delay. In this work we limit our study to the plasma in the lens.
The rotation measure 
\be
RM = \frac{r_e^2}{2\pi e} \int n_e(r) B_\parallel\, d z_l,
\ee
%
where $B_\parallel$ is the magnetic field along the l.o.s.
Due to the velocity difference between the L- and R-polarisation, there will be an arrival time difference during the propagation of distance $D$ in a uniform magnetic field $\vec{B}$,
\be
\Delta t_{LR} = \frac{2 \lambda^3 RM }{\pi c}.
\ee
Such a delay will cause a rotation of the linear polarisation angle by 
\be
\Delta \phi = \lambda^2 RM, \label{eq:faraday}
\ee
which is also known as the Faraday rotation. Thus one can use the variation of linear polarisation of the wavelength to estimate the magnetic field along the l.o.s.,  $\langle B_\parallel\rangle \sim RM/DM$. One approximation is that the deflection from plasma lensing is neglected, which will affect the Faraday rotation. 

First, we consider the transverse gradient of the magnetic field which does not correlate with electron density along the z-direction.
Then the lensing potential can be written as
\be
\psi = \theta_0^2 a(\theta) \pm \theta_b^2 a(\theta) b(\theta),
\ee
where we define
\begin{align}
\theta_0 &\equiv \lambda \rund{\frac{r_e N_0}{2\pi D_t}}^{1/2},\\
\theta_b &\equiv \rund{\frac{\lambda^2 N_0 r_e}{2\pi D_t}\dfrac{2\lambda r_e B_0}{e c \mu_0}}^{1/2}
=\theta_0 \rund{\dfrac{2\lambda r_e B_0}{e c \mu_0} }^{1/2} 
=\rund{\frac{\lambda^3 RM_0}{\pi D_t}}^{1/2},
\end{align}
where $D_t=D_dD_s/D_{ds}$.
The function $a(\theta)$ gives the profile of the projected electron density, i.e. $N_e=N_0\, a(\theta)$, and $b(\theta)$ gives the profile of the magnetic field, $B(\theta)=B_0 b(\theta)$.
The deflection angle can be given by
\be
\alpha_{L,R} = \theta_0^2 \nabla a(\theta) \pm \theta_b^2 \nabla(a(\theta) b(\theta))
\ee
The arrival time difference between the left and right polarisation can be given by 
\be
\Delta t_{LR}=\frac{(1+z)D_t}{c} \frac{\alpha_L^2 - \alpha_R^2}{2} + \frac{2 \lambda^3 RM}{\pi c},
\ee
where the second term is known as Faraday delay \citep[e.g.][]{1994ApJ...422..304T,2019MNRAS.484.5723L}. 
We will use $DM_0$ and $RM_0$ to indicate the dispersion measure and rotation measure at the centre of the lens. Then the phase rotation at position $\theta$ including the lensing effect is
\be
\Delta \phi = (1+z) \lambda^2 \theta_0^2 RM_0 \nabla a(\theta) \nabla (a(\theta)b(\theta))  + \lambda^2 RM_0 a(\theta)b(\theta).
\ee
The first term on the right-hand side is the ``geometric rotation'' $\Delta\phi_{geo}$, and the second term is the Faraday rotation $\Delta\phi_{RM}$. Since $\theta_0$ is proportional to $\lambda^2$, the geometric rotation has a different wavelength dependent than the Faraday rotation, i.e. $\Delta\phi_{geo}\propto\lambda^4$.
Usually, the geometric term is extremely smaller than the second one and can be safely neglected. In case of a high density of the electron, strong magnetic field, or large gradient of electron density, such a term can have a non-neglect contribution. The wavelength in the equation is that of the photon at the lens. And the RM is the observed one, which has an extra factor $1/(1+z)^2$ corresponding to the RM at redshift $z$. Thus both two delays are redshift dependent. For simplicity, we consider the low redshift case at current work, i.e. $z\sim0$.

\begin{table}
\begin{tabular}{c|c}
\hline
Symbol       &Description \\
\hline
$\theta$     & angular coordinate in the lens plane  \\
$\beta$     & angular coordinate in the source plane  \\
$\alpha$    & deflection angle\\
$\psi(\theta)$      & effective lens potential \\
$D_t$    & $=D_d D_s/D_{ds}$ \\
\hline
$n_e$    & 3D electron density\\
$\omega$, $\lambda$     & observed frequency, wavelength \\
$\omega_e$   &plasma frequency \\
$\omega_B$    &cyclotron frequency \\
\hline
$r_e$    & classical radius of the electron $r_e=\frac{e^2}{m_e c^2}$\\
$e$      & charge of the electron \\
$m_e$    & mass of the electron\\
\hline
\end{tabular}
\caption{Summary table of quantities and parameters used in this paper.}
\label{tab:defineparas}
\end{table}

First, we assume the magnetic field follows the same profile as the electron density in the rest of this work, i.e. $b(\theta)=a(\theta)$. We adopt a Gaussian profile for our plasma lensing,
\be
N_e(\theta) = N_0 {\rm exp}\rund{-\frac{\theta^2}{2\sigma^2}},
\label{eq:Gauss-positive}
\ee
where $\sigma$ is the width of the lens. The total rotation (geometry and Faraday) at image position $\theta$ is then
\be
\Delta \phi = \eck{2
\rund{\frac{\theta_0}{\sigma}}^2 \rund{\frac{\theta}{\sigma}}^2 {\rm exp}\rund{-\frac{3\theta^2}{2\sigma^2}}
+{\rm exp}\rund{-\frac{\theta^2}{\sigma^2}}} \lambda^2RM_0.
\label{eq:gauss-rotation}
\ee
The first term in the bracket presents the relative strength of the geometric rotation. The interesting points can be found here. First, we can see that the rotation strongly depends on the image position (the separation between the image and the centre of the lens) $\theta$. A similar situation can be found in the time delay of gravitational lensing. In the original paper of Shapiro \citep{1964PhRvL..13..789S}, it is stated that the time delay caused by the deflection, i.e. geometric delay is negligible compared with the potential delay. However, it is necessary to include such a contribution in the analysis of strong lensing time delay. The reason is where the image formed by the lens \citep{2020CQGra..37t5017T}. An image formed near the critical curve will experience a strong deflection and has a significant geometric time delay. A large geometric delay can also change the dispersion relation and bias the estimate of the electron density \citep[][]{er+2020}. A similar situation can be found here. If the image is formed near the lens, the geometric rotation is necessary to be taken into account. The other aspect is the property of the lens, or the density gradient of the lens, e.g. for Gaussian lens $\theta_0/\sigma$. Moreover, in the Gaussian profile, since $\theta_0^2/\sigma^2\propto N_0 D_{ds} D_d /D_s$ (since $\sigma\propto 1/D_d$), besides the projected electron density $N_0$, the distances of the lens and the source are important as well. At a cosmological distance, such as FRBs, the geometric term will become important. For a given distance of the source, the factor $D_{ds} D_d/D_s$ reaches a maximum when $D_d=D_{ds}$, i.e. when the lens locates at the middle point between the source and us. The plasma lens in the host galaxy or in the Milky Way will play a relatively small effect. Moreover, from Eq.\,\ref{eq:gauss-rotation} we can see the ratio between the geometric to Faraday rotation does not depend on the strength of the magnetic field at all, but only on the gradient of the electron density. This is because we assume that the magnetic field has the same profile as that of electron density.

The extra deflection by the RM can split the caustics of the lensing. Without RM, the caustics of the Gaussian model lens are generated by the radial magnification \citep{er&rogers18}. The magnetic field, i.e. RM will cause two effects: first, it will generate a caustic of tangential magnification, and only to one polarisation at a specific frequency. However, it requires strong RM, i.e. $\theta_b\sim \sigma$. Since $\theta_b$ is frequency dependent, the caustic has a frequency shift \citep{2019MNRAS.484.5723L}. The second effect is that it will split the caustic by radial magnification slightly, which also requires a large RM or $\theta_b$. Both of them will cause a difference between the flux of two polarisation modes, which will be manifested by a toy example later.

If the density of the free electron is a power law profile, $\theta_0$ and $\theta_b$ will be different. We start with the three dimension density profile
\be
n_e(r) = n_0 \rund{\frac{R_0}{r}}^h, \qquad h>0,
\ee
where $n_0$ is a constant representing the electron density at radius $r=R_0$. We consider a case of $h=2$ for the electron density and $h=1$ for the magnetic field, i.e. $B(r)=B_0(R_0/r)$. The deflection angle is
\be
\alpha (\theta) = \frac{\theta_0^3}{\theta^2} \pm \frac{\theta_b^4}{\theta^3},
\ee
where the characteristic radius is \citep{2010MNRAS.404.1790B,BTreview15}
\be
\theta_0 \equiv \rund{\lambda^2\dfrac{D_{ds}}{D_sD_d^2} \dfrac{r_e n_0 R_0^2}{\sqrt{\pi}} \dfrac{\Gamma(3/2)}{\Gamma(1)}}^{1/3},
\label{eq:t0-ne}
\ee
and 
\be
\theta_b \equiv \rund{2\lambda^3\dfrac{D_{ds}}{D_sD_d^3} \dfrac{r_e^2 R_0^3}{\sqrt{\pi}}RM_0 \dfrac{\Gamma(2)}{\Gamma(3/2)}}^{1/4}.
\label{eq:tb-ne}
\ee
Similarly, the rotation by the magnetic field will be increased due to geometric rotation. The lens properties are slightly different, and different profiles of the magnetic field can cause complicated splitting of the two circular modes.

\section{Simulated examples}
\label{sec:example}

We adopt toy models to compare the intensity between L- and R-mode and the rotation to the polarisation. We take a positive Gaussian model and an under-dense model. The free electron density has a wide range in the Milky Way \citep[e.g.][]{NE2001,2017ApJ...835...29Y}. RM can also have large variation, e.g. $RM\sim 10-200$ rad\,m$^{-2}$ \citep[][]{1981ApJS...45...97S,2009ApJ...702.1230T}. We adopt a conservative value for the central electron density and magnetic field $N_0=120$ pc\,cm$^{-3}$, $RM_0=50$ rad\,m$^{-2}$ for a Gaussian profile, and $\sigma=10$ AU ($\sim$ 0.026 arcsec).
The distance of the lens and source are taken from a pulsar plasma lensing event PSR B0834+06, $D_d=389$ pc, $D_s=620$ pc \citep{2016MNRAS.458.1289L,2022arXiv220806884Z}.
%

\begin{figure}
\includegraphics[width=6cm]{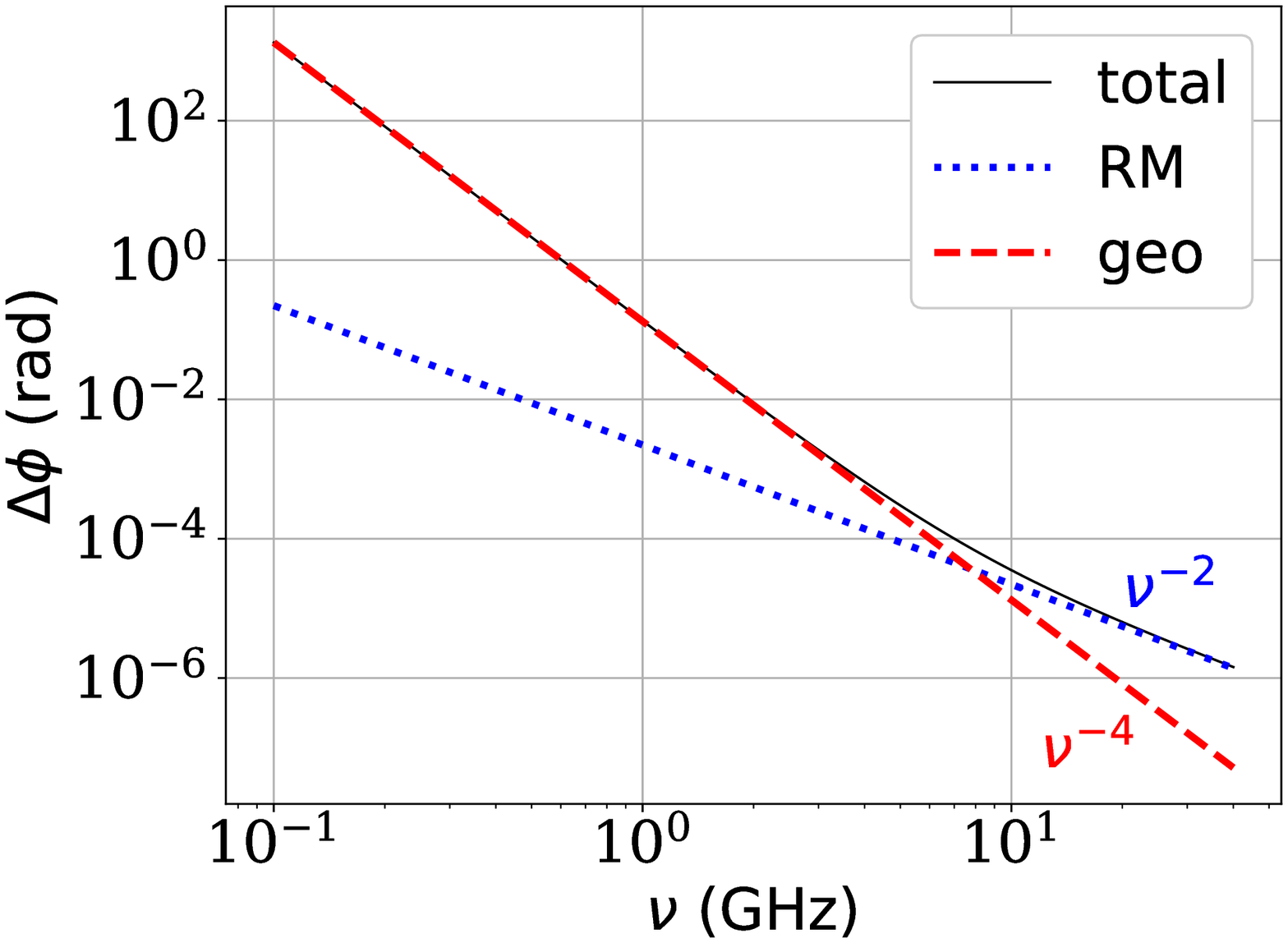}
\includegraphics[width=6cm]{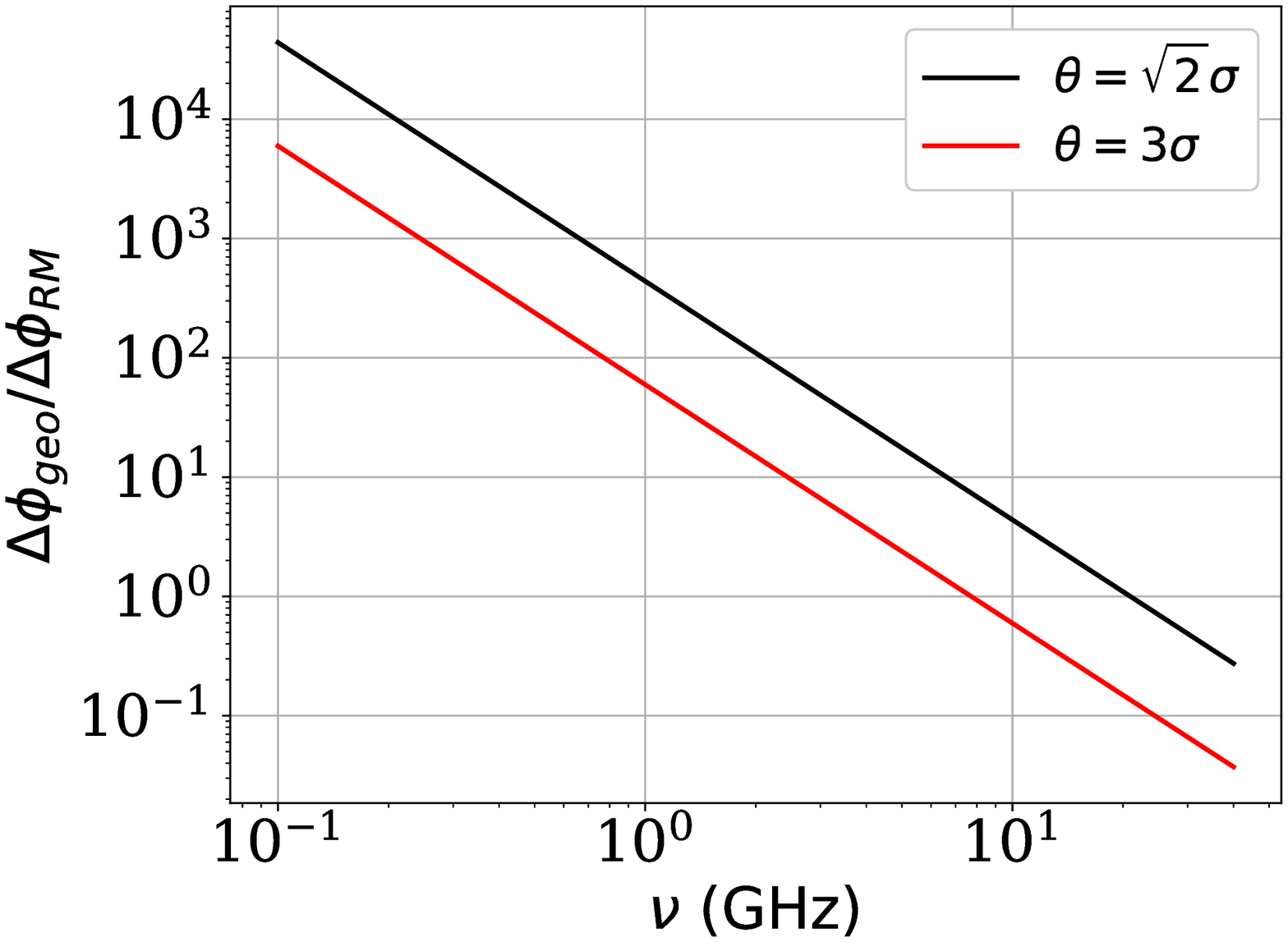}
\includegraphics[width=7cm]{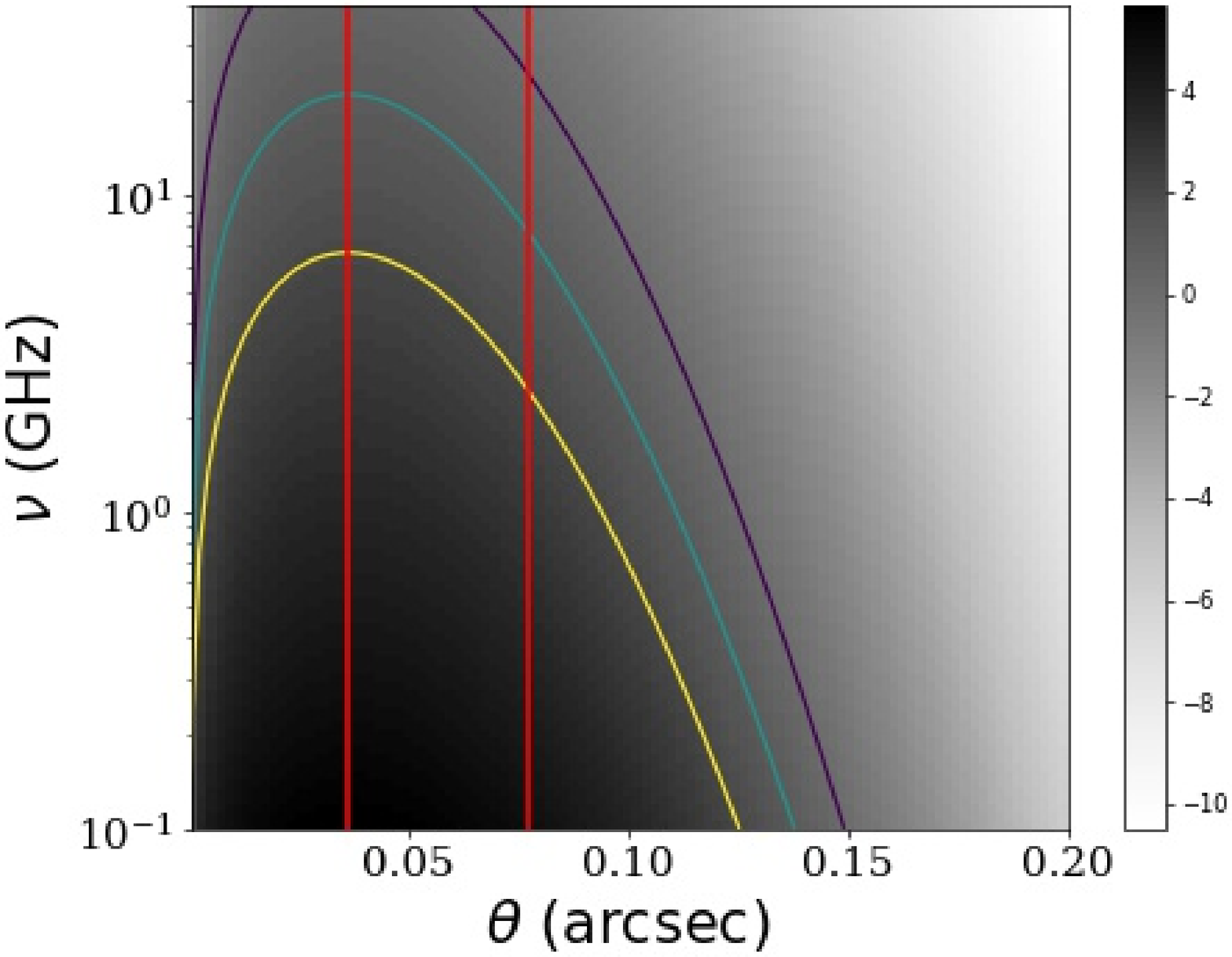}
\caption{$Top-$ the geometric rotation (red dash line), the Faraday rotation (blue dotted line), and total rotation (black solid line), at $\theta=3\sigma$. $Middle-$ ratio between geometric rotation to Faraday rotation as a function of frequency at two image positions. $Bottom-$ ratio of rotation at different image positions and frequencies. The dark (light) colour stands for a high (low) ratio. The colour bar gives the log-scale. The vertical lines indicate two image positions, $\theta=\sqrt{2}\sigma,\, 3\sigma$ respectively. The purple, cyan, and yellow curve indicates the contour where the ratio equals $0.1, 1, 10$ respectively.  }
\label{fig:gauss-rotation}
\end{figure}
We first adopt a positive Gaussian density model (Eq.\,\ref{eq:Gauss-positive}), and compare the Faraday rotation and the geometric rotation (Fig.\,\ref{fig:gauss-rotation}). In the first panel, we present the Faraday rotation and geometric rotation at $\theta=3\sigma$. They follow different power laws on the frequency, -2 and -4 respectively. They cross at about $\nu\sim 8$ GHz. Above this frequency, the geometric rotation can be neglected. While below that, the estimate of the magnetic field will be problematic if one only considers the Faraday rotation. We present the ratio between the geometric rotation to Faraday rotation at two image positions. One is at where the ratio reaches a maximum in the Gaussian model, i.e. $\theta=\sqrt{2}\sigma$. The other one is at $\theta=3\sigma$. The last panel shows the ratio between geometric rotation to Faraday rotation by the grey shadow, in logarithmic scale. Within the cyan curve, i.e. at low frequency and image near the lens, the geometric rotation will dominate. Out of the purple curve, the geometric rotation only causes a small ``perturbation''. When the image is close to the lens, the geometric term dominates. However, a concentrate plasma lens is diverging, i.e. when the lens and source are well aligned, it causes strong de-magnification. We have a low chance to observe an image near the centre of the lens.
We thus define an average ratio $r$ between the geometric to the Faraday rotation with magnification as weighting by 
\be
\bar r \equiv \dfrac{\int \int_0^{\beta_c} \d^2 \beta\,  \mu(\beta)\, r(\beta) }{\pi \beta_c^2},
\label{eq:mean-ratio}
\ee
where $\beta_c$ is the cross-section of the lens, which we take roughly $4\beta_0$ here. Different values of $\beta_c$ have been used, and no dramatic change has been found. The average ratio can vary from about one hundred at low frequency ($<1$ GHz) to $\sim 10^{-3}$ at $\sim 10$ GHz in the positive Gaussian model.

\begin{figure}
\centerline{\includegraphics[width=9cm]{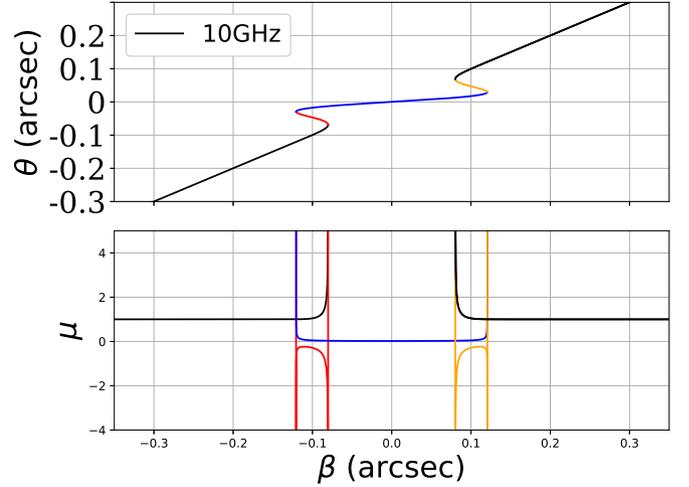}}
\caption{The $\beta-\theta$ relation (top), and the magnification curves (bottom) for a Gaussian plasma lens at 10 GHz. Different colours are used for different source positions. }
\label{fig:gaussian_young}
\end{figure}

We move a source behind the lens from $\beta=-0.3$ arcsec to $\beta=0.3$ arcsec along the axis of the plasma lens. In Figs.\ref{fig:gaussian_young}, we show the image-source position and the magnification curve on the source plane. Although at small $\theta$, $\mu$ approaches zero, the two critical curves are still within the region of the purple curve in Fig.\,\ref{fig:gauss-rotation}. Therefore, when the image is near the critical curves, i.e. strongly magnified, there is a large probability to detect the lensed image, and with a large geometric rotation.

\begin{figure*}
\centerline{\includegraphics[width=8cm]{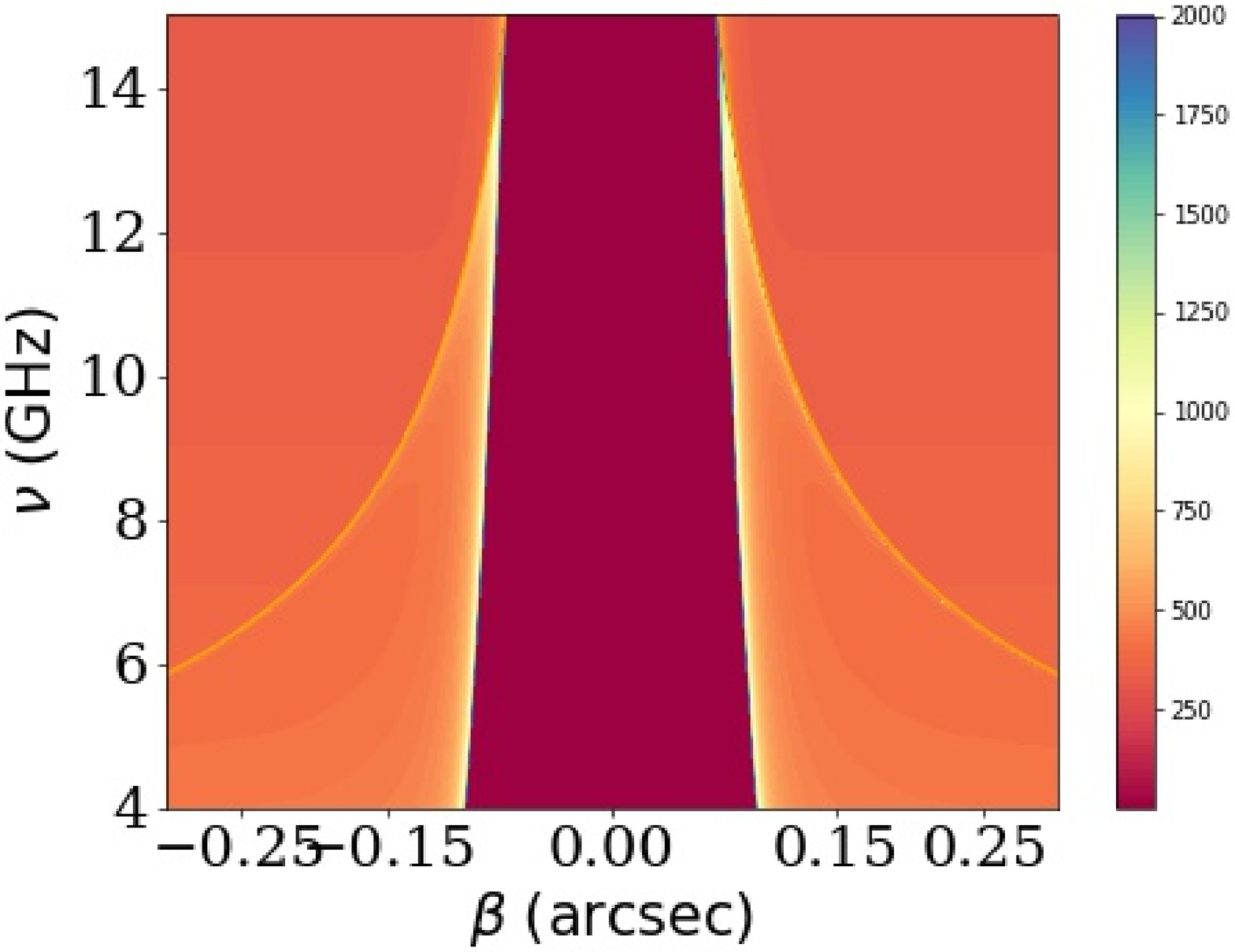}
\includegraphics[width=8cm]{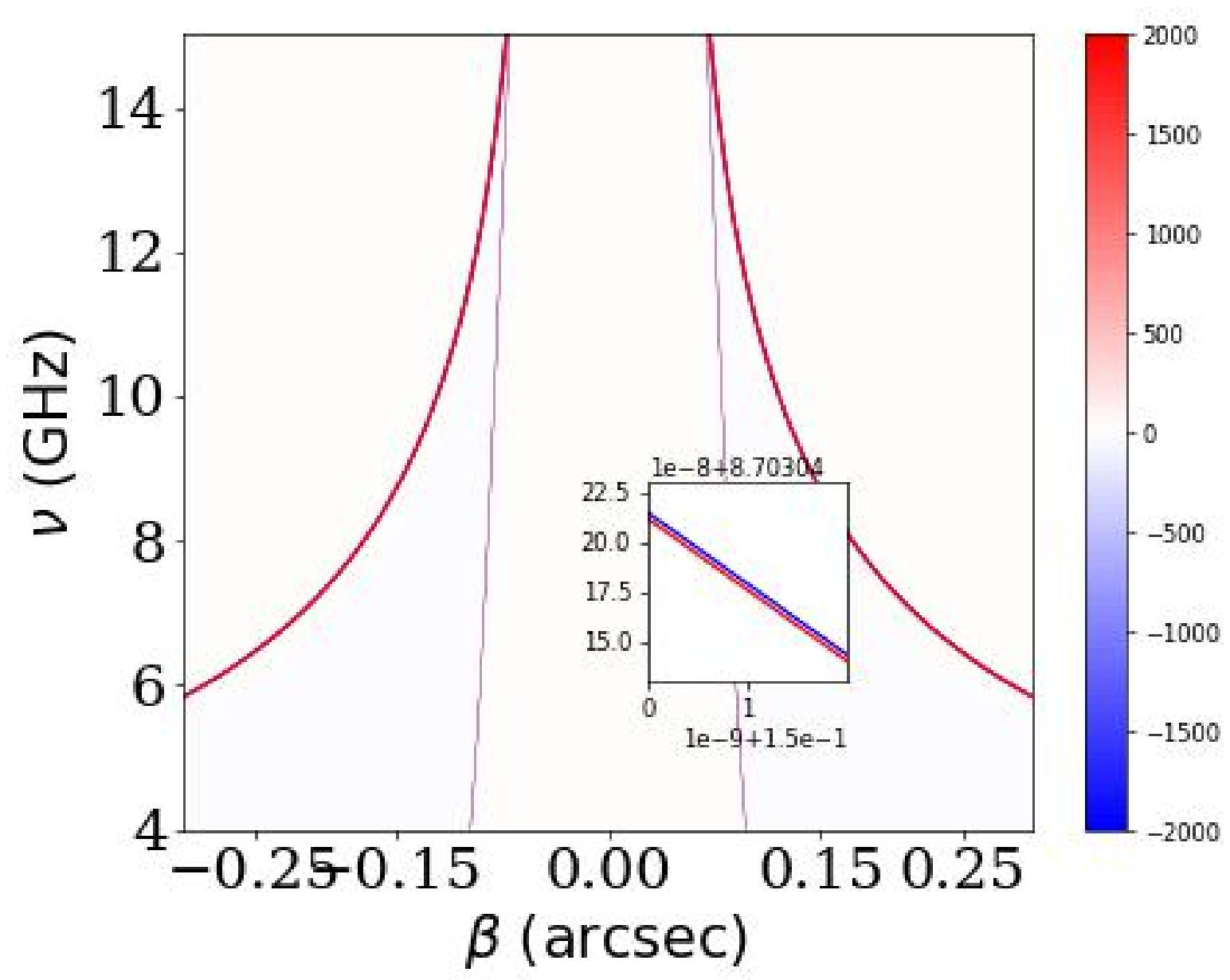}}
\caption{The simulated flux density (in mJy) on frequency-source position plane of the lensed images. The intensity is truncated at 2000 mJy for better visibility. $Left-$ (Stokes-I), $Right-$ Stokes-V. The spatial separation of the two caustics near 10 GHz is about $10^{-10}$ arcsec (zoom-in panel). }
\label{fig:freq-time}
\end{figure*}

We simulate the intrinsic flux of the source by
\be
F(\nu)=F_0\rund{\frac{\nu}{\nu_0}}^{-0.22},
\ee
with $F_0=176$ mJy and $\nu_0=10$ GHz \citep{2016ApJ...817..176T}. We assume that the source is partially linear polarised, but as we can see later that both linear and unpolarised modes will be changed. The intensity of the lensed image at various frequencies is shown in Fig.\,\ref{fig:freq-time}. When there are multiple images, the intensity is the sum of all the lensed images. Within the inner caustics, no images are formed, resulting in zero intensity. Near the caustics, the yellow and blue colours represent the strong magnification of the images near the critical curves. {The right panel depicts the difference between the left- and right polarisation, i.e. a Stokes-V component. As we use a partial linear mode of the source, there is no signal detected most of the time. However, in an extremely narrow region near the critical curve, we detect a significant Stokes-V component. The position of caustics for left- and right-modes are slightly different in reality. Therefore, the two circular modes will be immensely magnified at slightly different positions and left with only one mode to be detected at one caustic, regardless of the initial polarisation mode of the source. At $\sim10$ GHz, the separation between the caustics of two modes is about $10^{-10}$ arcsec.
Assuming the relative speed of the lens is $100\,km/s$, the circular mode will switch within 0.1 seconds. However, for a lens at 5 kpc, the Fresnel scale at 10 GHz is $\sim10^{-6}$ arcsec. Thus the image size will be larger than the magnified event, or the split of the caustics. It will be difficult to measure the split, which is much smaller than the image size. 
At lower frequencies, the split and the switching time become bigger. For example, the circular mode can remain for $\sim 1$ second at $\sim 3$ GHz. The Gaussian lens can generate two sets of caustics at lower frequencies. The same phenomena happen to the inner caustic as well, but it is less prominent because the split of the caustic is weaker.} Moreover, near the critical curve, the images will be highly magnified, which increases the probability of detection.

\begin{figure}
\includegraphics[width=7cm]{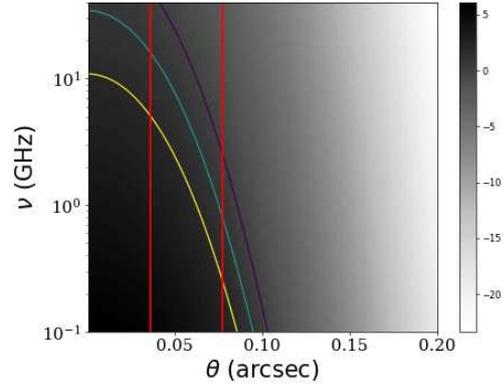}
\caption{Same as bottom panel of Fig.\,\ref{fig:gauss-rotation} but for an under dense Gaussian density profile.}
\label{fig:gauss-neg-rotation}
\end{figure}

In the second test, we adopt an under-dense Gaussian model of electron density which is a convergent lens \citep[e.g.][]{2012MNRAS.421L.132P},
\be
N_e=N_0 - N_0 {\rm exp}\rund{-\dfrac{\theta^2}{2\sigma^2}},
\ee
or in our way $a(\theta)=b(\theta)=(1-{\rm exp}(-\theta^2/(2\sigma^2)))$.
The deflection angle will be 
\be
\alpha_{L,R} = \frac{\theta_0^2 \theta}{\sigma^2} g(\theta) \pm \frac{2\theta_b^2 \theta}{\sigma^2} g(\theta) (1-g(\theta)),
\ee
where $g(\theta)={\rm exp}(-\theta^2/(2\sigma^2))$.
The total rotation by the plasma lens with a magnetic field will be 
\begin{align}
\Delta \phi &=\lambda^2RM_0 \nonumber \\
&\eck{2\rund{\frac{\theta_0}{\sigma^2}}^2 \rund{\frac{\theta^2}{\sigma^2}}{\rm e}^{-\frac{\theta^2}{\sigma^2}}\rund{1-g(\theta)}+ \rund{1-g(\theta)}^2}.
\end{align}
In Fig.\,\ref{fig:gauss-neg-rotation}, we show the ratio between the geometric rotation to the Faraday rotation. A similar situation can be found, i.e. when the image is formed near the lens, the geometric rotation dominates. However, since the under-dense plasma lens is a converging lens, the images within the critical will not be de-magnified. The probability of detecting images with high magnification, or the cross-section of high magnification is higher than that of the diverging lens. 

\begin{figure}
\includegraphics[width=8cm]{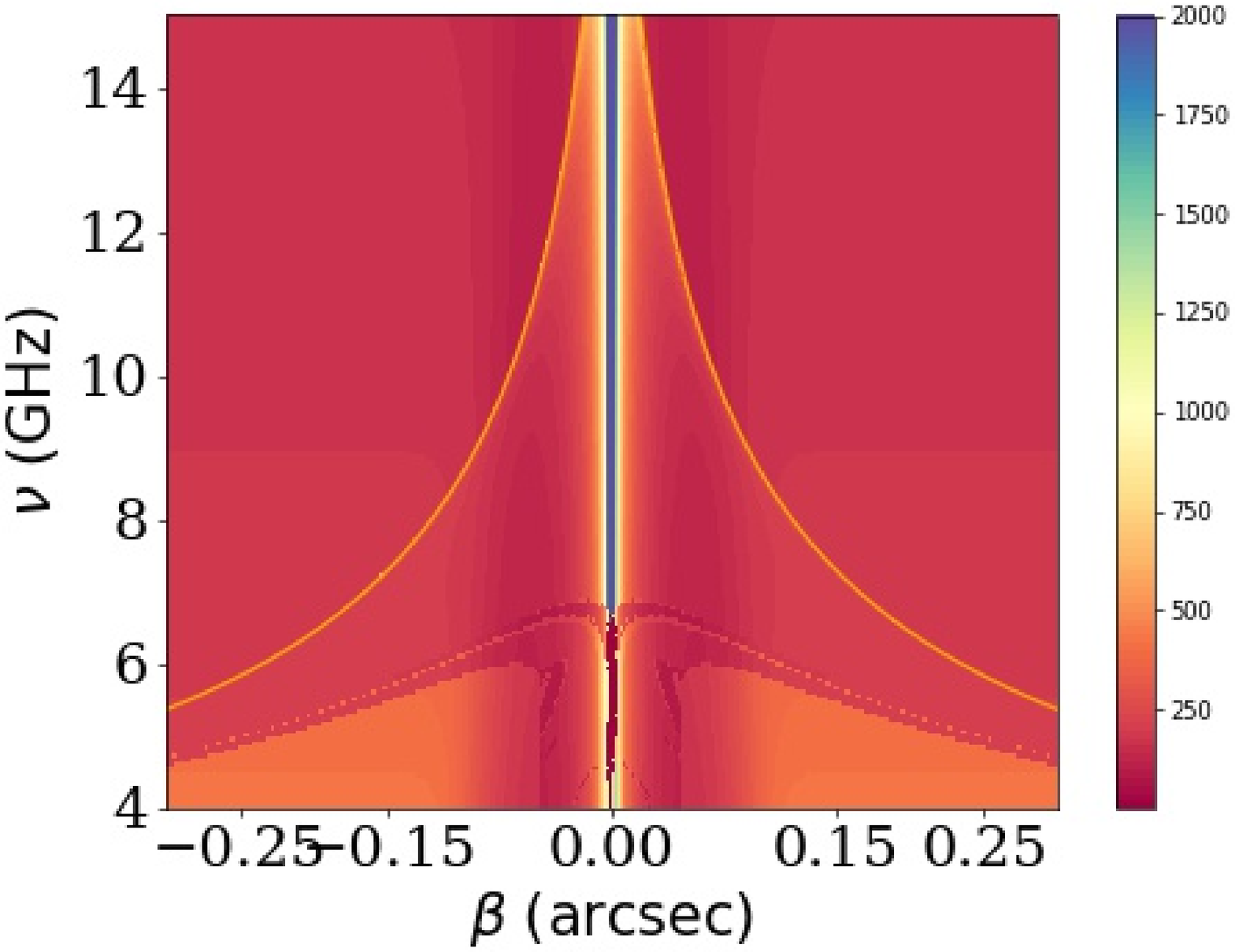}
\includegraphics[width=8cm]{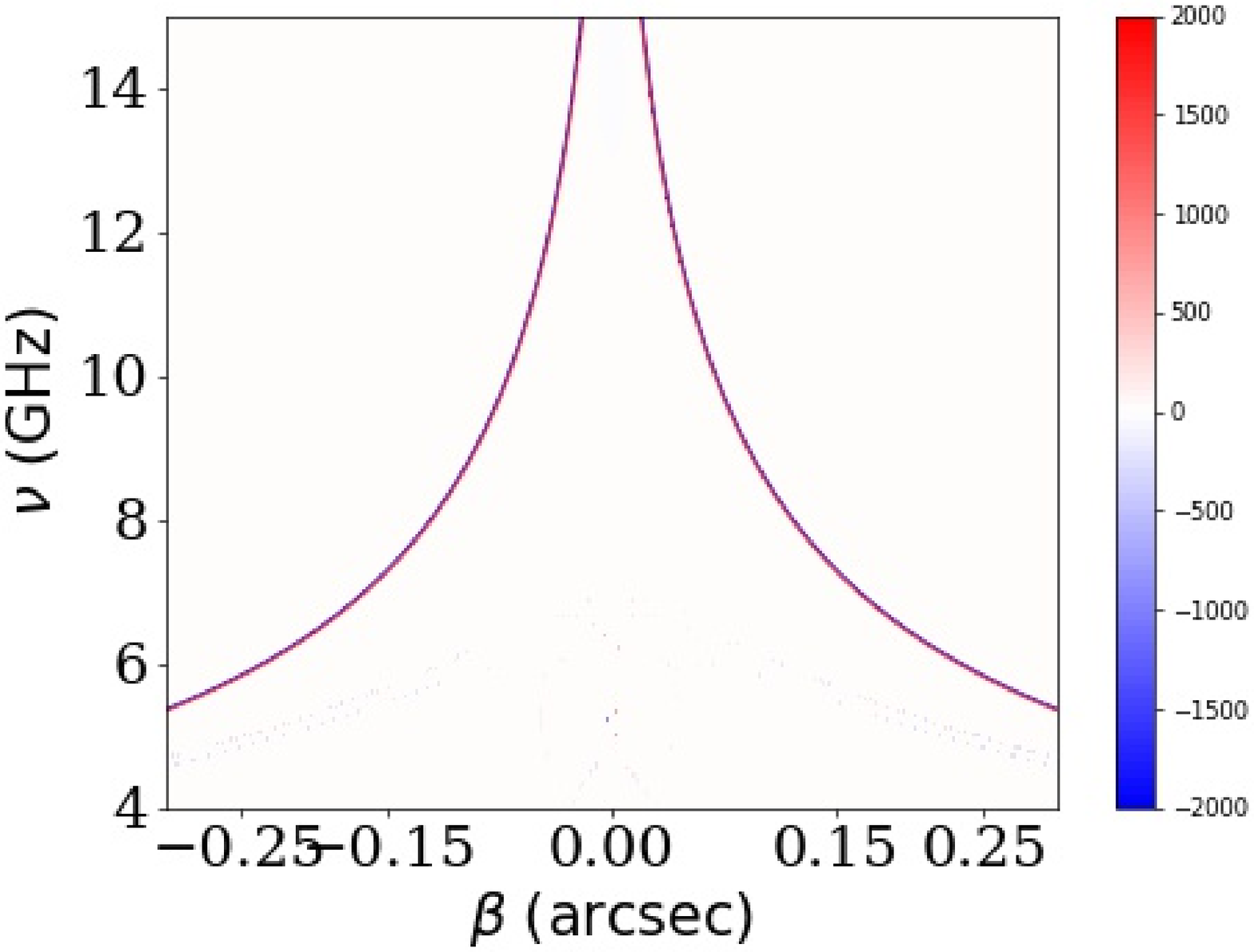}
\caption{Same as Fig.\,\ref{fig:freq-time} but for an under dense Gaussian density profile. The top is for Stokes-I and the bottom is for Stokes-V. }
\label{fig:freq-time-neg}
\end{figure}

In Fig.\,\ref{fig:freq-time-neg}, we adopt the same intrinsic flux and show the lensed image flux. Unlike the diverging lens, there is no de-magnification region. Near the out critical curve, the lens induces circular polarisation as well. In the under-dense lens, there is a large frequency range to detect the difference between the two polarisation modes.
We estimate the average ratio between the geometric rotation to the Faraday rotation using Eq.\,\ref{eq:mean-ratio} for the under-dense model as well. In most cases of our interest (images near the lens, and lower than $\sim 30$ GHz), the average ratio is greater than one. The reason is that the lensing magnification boosts the probability to detect the images with high geometric delay. Therefore, in the under-dense lens model, geometric rotation is common and can dominate the total rotation in many cases. Thus it is necessary to take into account in estimate of the magnetic field.

\subsection{Differential arrival times}
The dispersion relation gives the inverse quadratic $t_{\rm DM} \propto \nu^{-2}$DM delay of pulsars and FRBs, which is caused by a constant distribution of plasma. Besides that, RM, the gradient of the electron density, and the gradient of RM will cause extra delays as well. In the previous analysis, the higher-order terms are concluded to be negligible \citep[e.g.][]{1992ApJ...385..273P}. Recent observations find large value RM and strong variations of DM, e.g. in the repeating FRB 121102, RM has been found to be $\sim 10^5$ rad\,m$^{-2}$\citep{2018Natur.553..182M,2018ApJ...868L...4M}. Thus in this section, we estimate the effects of RM in some extreme cases.

\begin{figure}
\centerline{\includegraphics[width=8cm]{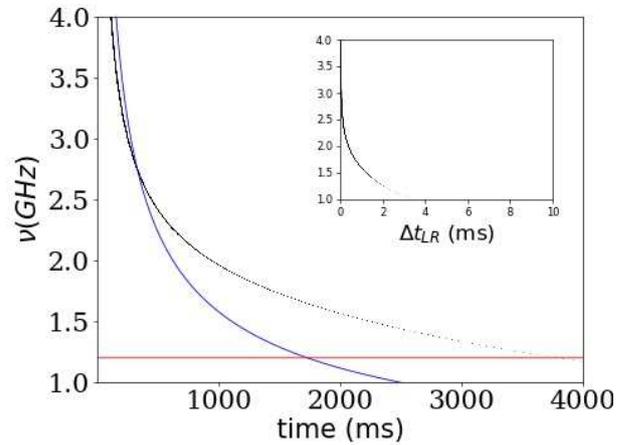}}
\caption{The simulated radio dispersion signal. The blue curve presents the dispersion relation for a constant DM at the image position ($\theta\approx 0.03$ arcsec at $\nu=1$ GHz). The grey shadow shows the total time delay by the plasma lensing for left- and right polarisation. Below the red horizontal line, the lens becomes super-critical. But only the primary image is shown in this figure. In the inset panel, the grey scale shows the arrival time difference between the left and right polarisation.}
\label{fig:td-nu}
\end{figure}

We adopt a mock example in this section. A radio pulse source with a width of 0.2 milli-second behind a Gaussian plasma lens is used (Eq.\,\ref{eq:Gauss-positive}). In order to manifest the effect of RM, we use an extremely large value, $RM_0=5\times10^7$ rad\,m$^{-2}$, but a relatively small $N_0=1000$ pc\,cm$^{-3}$. The size of the lens is $\sigma=1$ arcsec, which makes the lens sub-critical at high frequency ($\sim 1.2$ GHz), i.e. no critical curve or multiple images, and super-critical at low frequency, $\nu < 1$ GHz. We align the source to the centre of the lens by $\beta=1.5$ arcsec, inside the caustic at low frequency. 
The frequency delay relation is shown in Fig.\,\ref{fig:td-nu}. Since the image position will depend on the frequency, at the image position, the electron density, and rotation measure will vary between $N_e(\theta)\approx350-930$ pc\,cm$^{-3}$, $RM(\theta)\approx6.5\times10^{6}-4.3\times10^{7}$ rad\,m$^{-2}$ respectively. The blue curve presents the delay with a constant DM$=600$ pc$cm^{-3}$. The grey shadow shows the full lensing delay. They almost overlap at high frequencies and the lensing delay is significantly larger than the blue curve at low frequencies. The grey scale indicates the flux of the lensed image. It becomes faint at low frequency due to the plasma lensing de-magnification. On top of that, we show the different arrival times between the left and right polarisation in the small panel, which can reach a few milliseconds at low frequencies. Such a difference will either split the signal into two pulses or broaden the width of the signal.

The splitting of left and right polarisation can be observed with high RM. For example, in case of $\Delta t_{L,R}=1$ milli-second at 1 GHz, RM has to be $\sim 10^7$ rad\,m$^{-2}$ \citep[e.g.][]{2019ApJ...870...29S}. The contribution of the geometric delay has not been included, which can enhance the splitting with a density gradient. As every coin has two sides, the diverging plasma lensing will de-magnify the image, and thus may reduce the probability of the detection of such kinds of images. The exact situation can be complicated depending on the density profile of the electrons and the magnetic field. A conservative estimate with a Gaussian model, e.g. weak de-magnification or even magnification, will be around $\theta\sim \theta_0 \sim \sigma$, and $\Delta t_{\rm geo}/\Delta t_{RM}\sim 1$, i.e. the geometric delay can double the splitting. In our estimate of the ratio (e.g. Eq.\,\ref{eq:mean-ratio}, Fig.\,\ref{fig:gauss-rotation}), the geometric term can be greater than the RM term to $\sim 10^3$ at low frequency, a few hundred MHz. In other words, a magnetic field of $RM \sim 10^4$ rad\,m$^{-2}$ has the possibility to split the signal. It is even more interesting in the under-dense model since the magnification effect will enhance the probability of detection. The condition is that the image formed within the lensing cross-section. The exact enhancement by the geometric delay needs further sophisticated studies. But in the under-dense lens with $RM\sim 10^5$ rad\,m$^{-2}$, we expect to observe two close-up pulses with left- and right polarisation modes or a large width of one mixed pulse signal.

\section{Discussion and summary}
\label{sec:summary}
Plasma lensing occurs when the density gradient in plasma deflects a light ray that is propagating through the ionised medium. In magnetic media, radio signals exhibit birefringence, which is usually measured through the Faraday rotation of linearly polarised radiation. This happens because the refractive index of the two circular polarisation modes is different. The slight difference in the propagation of the two modes causes an additional arrival time difference that can lead to the rotation of linear polarisation. This geometric delay is  inversely proportional to frequency to the power of $4$, and it strongly depends on the density gradient of the plasma. This effect is different from the Faraday rotation. 

In this study, we compare the rotation of linear polarisation due to geometric effects and Faraday rotation for Gaussian density and under-dense Gaussian density models. Our findings show that the geometric delay can significantly increase the polarisation rotation, especially in the under-dense model. Therefore, it is necessary to consider the geometric effect when estimating the magnetic field using Faraday rotation. 

The birefringence can split the two circular polarisation modes. In the case of an extremely large magnetic media, there is a possibility to see two pulses of the signal or large width of the pulse. With the addition of geometric delay, a split of 1 milli-second at 1 GHz will require RM$\sim10^{5}$ rad\,m$^{-2}$. This value has been found in our universe, such as FRB 121102. Therefore, we expect that the lens can change the shape of the pulse, especially at low frequencies. 
We adopt a toy model for the intrinsic flux of a radio source and simulate the lensed flux for observation. Rapid flux variation occurs near the critical curve of the lensing. Moreover, due to the split of lensing caustic, we expect that the lens can induce Stokes-V mode and a strong variation of the polarisation modes in an extremely narrow region near the critical curve, regardless of the initial polarisation of the sources. The split can be smaller than the spatial size of the source, and is difficult to detect. The possibility will depend on the magnetic field and the frequency of observation.

Our study is a simple representation and does not fully describe the real observations. The plasma density model that we adopt only considers the profiles on a relatively large scale, such as from hundreds of AU to kpc. Small-scale fluctuations, which have been observed within the solar system \citep[e.g.][]{2019NatAs...3..154L}, will increase the multi-path scattering that broadens pulses. The split pulses can merge due to the broadening width. In the end, both scattering and birefringence make the estimate of the intrinsic width of the pulse difficult.
The Gaussian density model is widely used in plasma lensing due to its analytic simplicity. More realistic density models are necessary, either from simulations or better observations. Especially the multi-component density profiles on different scales worth more detailed investigations. For cosmological sources, such as FRBs, the multiplane lens will induce different redshift dependence and require further studies.

\section*{Acknowledgements}
We would like to thank the referee for a high evaluation of our work and suggestions that improved the clarity of our manuscript. 
XE is supported by the NSFC Grant No. 11873006, 11933002, and the China Manned Space Project with No.CMS-CSST-2021-A01, No.CMS-CSST-2021-A07, No.CMS-CSST-2021-A12. XS is supported by the National SKA program of China (2022SKA0120101).

\section*{Data Availability}
The data underlying this article will be shared on reasonable request to the corresponding author.
\bibliographystyle{mnras}
\bibliography{bfield,lens,plasmalens}

\bsp	
\label{lastpage}
\end{document}